\documentclass[
    a4paper,
    reprint,
    onecolumn,
    notitlepage,
    oneside,
    amsfonts,
    amssymb,
    amsmath
]{revtex4-2}
\usepackage{placeins}
\usepackage{graphicx}
\usepackage{dcolumn}
\usepackage{bm}
\usepackage{multirow} 
\usepackage{caption}
\usepackage{amsmath}
\usepackage{hyperref}

\usepackage{float}
\usepackage{scalerel}
\usepackage{tikz}
\usetikzlibrary{svg.path}

\definecolor{orcidlogocol}{HTML}{A6CE39}
\tikzset{
  orcidlogo/.pic={
    \fill[orcidlogocol] svg{M256,128c0,70.7-57.3,128-128,128C57.3,256,0,198.7,0,128C0,57.3,57.3,0,128,0C198.7,0,256,57.3,256,128z};
    \fill[white] svg{M86.3,186.2H70.9V79.1h15.4v48.4V186.2z}
                 svg{M108.9,79.1h41.6c39.6,0,57,28.3,57,53.6c0,27.5-21.5,53.6-56.8,53.6h-41.8V79.1z M124.3,172.4h24.5c34.9,0,42.9-26.5,42.9-39.7c0-21.5-13.7-39.7-43.7-39.7h-23.7V172.4z}
                 svg{M88.7,56.8c0,5.5-4.5,10.1-10.1,10.1c-5.6,0-10.1-4.6-10.1-10.1c0-5.6,4.5-10.1,10.1-10.1C84.2,46.7,88.7,51.3,88.7,56.8z};
  }
}

\newcommand\orcidicon[1]{\href{https://orcid.org/#1}{\mbox{\scalerel*{
\begin{tikzpicture}[yscale=-1,transform shape]
\pic{orcidlogo};
\end{tikzpicture}
}{|}}}}

\begin{document}

\title{
\textsc{CosmoFit}: A Graphical User Interface for Bayesian Cosmological
Parameter Estimation with User-Defined \(H(z)\) Models
}

\author{Giovany Cruz\orcidicon{0000-0002-2877-6922}}
\email{giocruz@uv.mx }

\begin{abstract}
We present CosmoFit, a desktop graphical interface for Bayesian parameter inference in background cosmology. The application is designed to simplify cosmological analyses by reducing the need to manually program each step, while maintaining the transparency and reproducibility of the results. Users can define an expansion history \(H(z)\), specify fixed and sampled parameters, assign prior ranges and proposal scales, select the available observational likelihoods, and run Markov chain Monte Carlo analyses from a unified graphical interface. CosmoFit uses Cobaya for Bayesian sampling and GetDist for posterior analysis and visualization. The current version supports cosmic-chronometer data and the official Pantheon+, Pantheon+SH0ES, and Union3 supernova likelihoods available through Cobaya. For each run, the application stores the configuration files, metadata, execution logs, chains, and numerical summaries required to inspect and reproduce the analysis. The performance of CosmoFit was evaluated through unit tests, a parameter-recovery test using synthetic data, and an independent comparison with an analysis performed directly in Cobaya. The posterior mean values obtained with both procedures agree to within less than one percent. In addition, the synthetic-data test recovers the fiducial parameters within the expected credible regions. As an illustrative example, we perform a spatially flat \(w\)CDM analysis using Pantheon+SH0ES and obtain marginalized constraints and triangle plots directly from the interface. Within its current scope, CosmoFit provides an accessible, transparent, and reproducible tool for research and academic applications involving explicit models of the background cosmological expansion.
\end{abstract}

\maketitle


\section{Introduction}
\label{sec:introduction}

The increasing precision and diversity of cosmological observations have made statistical inference an essential component of modern cosmology. Measurements obtained from Type Ia supernovae, cosmic chronometers, baryon acoustic oscillations, the cosmic microwave background, and the large-scale distribution of matter are routinely used to constrain cosmological parameters and test competing descriptions of the expansion history of the Universe. In this context, Bayesian inference and Markov chain Monte Carlo methods provide a systematic framework for estimating model parameters, quantifying statistical uncertainties and parameter degeneracies, and evaluating the compatibility between theoretical predictions and observational data \cite{Trotta2008,LewisBridle2002}.

Several mature computational packages have been developed to support cosmological parameter estimation. Codes such as \textsc{CosmoMC} \cite{LewisBridle2002}, \textsc{MontePython} \cite{Audren2013,Brinckmann2019}, and \textsc{Cobaya} \cite{TorradoLewis2021} provide flexible environments for defining cosmological models, observational likelihoods, prior distributions, and sampling strategies. In particular, \textsc{Cobaya} offers a modular framework capable of combining theoretical calculations, external likelihoods, and different sampling algorithms through configuration-based analyses. These tools are extensively used in cosmology because they separate the statistical machinery from the physical implementation of the model and facilitate the construction of reproducible inference pipelines.

Despite these advantages, the direct use of general-purpose inference frameworks usually requires familiarity with programming languages, package management, command-line tools, configuration files, likelihood implementations, and posterior-processing utilities. For experienced users, this flexibility constitutes one of their main strengths. However, it may also represent an important entry barrier for students, educators, and researchers who wish to test a background cosmological model without first implementing a complete computational pipeline. This situation is particularly relevant for models whose primary theoretical prediction is a modified expansion law ($H(z)$), while their observational analysis relies on standard background probes such as Type Ia supernovae or cosmic-chronometer measurements.

A graphical interface can reduce this technical barrier, but a scientifically useful interface must do more than hide command-line instructions. It must preserve the physical and statistical assumptions of the analysis, validate the mathematical model before sampling, record the exact configuration passed to the inference engine, isolate long numerical calculations from the graphical process, and retain the files required to reproduce the results independently. Furthermore, allowing users to introduce mathematical expressions presents an additional software-design challenge. A naive implementation based on the unrestricted evaluation of user-provided text could allow arbitrary code execution and compromise both security and reproducibility. A reliable application therefore requires a restricted and explicitly validated mathematical-expression system.

In this work, we present \textsc{CosmoFit}, an open-source desktop graphical interface for Bayesian parameter inference in homogeneous and isotropic background cosmology. \textsc{CosmoFit} allows users to define a cosmological expansion history through a mathematical expression for ($H(z)$), declare fixed and sampled parameters, assign prior intervals, select a supported observational dataset, execute a Markov chain Monte Carlo analysis, and inspect the resulting posterior constraints. The numerical inference is performed by \textsc{Cobaya}, whereas \textsc{GetDist} \cite{Lewis2019GetDist} is used to obtain marginalized parameter summaries and one- and two-dimensional posterior distributions. \textsc{CosmoFit} is therefore not intended to replace these established packages. Instead, it provides an accessible, transparent, and reproducible layer for configuring and executing background-level cosmological analyses.

A central feature of \textsc{CosmoFit} is its restricted parser for user-defined expressions. Each expression is parsed and validated before an inference run is created. It can also be evaluated numerically over a selected redshift interval in order to identify undefined domains, singularities, complex values, or non-finite predictions before the sampling stage. This approach provides flexibility in the definition of cosmological models while avoiding unrestricted execution of user-supplied instructions.

The graphical configuration is translated automatically into a valid \textsc{Cobaya} input. The resulting sampling process is launched in an isolated worker process, preventing long calculations or numerical failures from blocking the graphical event loop. Each analysis is stored in a dedicated output directory containing the original user input, a normalized internal configuration, the generated \textsc{Cobaya} configuration, runtime metadata, execution status, logs, posterior chains, and numerical summaries. Consequently, a run initiated from the graphical interface can be inspected, repeated, or reconstructed independently outside the application.

The current release of \textsc{CosmoFit} focuses on late-time background cosmology and assumes spatially flat distance relations. It supports direct measurements of ($H(z)$) from cosmic chronometers and the official \textsc{Cobaya} likelihood components associated with the Pantheon+, Pantheon+SH0ES, and Union3 Type Ia supernova compilations \cite{brout2022pantheon+,riess2022comprehensive,Rubin2025Union3}. Cosmological perturbations, Boltzmann-code calculations, and structure-growth observables are outside the scope of the initial version. This restricted scope is intentional: the main objective is to provide a reliable environment for testing standard and non-standard late-time expansion histories defined directly through ($H(z)$).

The scientific reliability of the graphical workflow is evaluated through several complementary verification and validation procedures. These include unit and integration tests of the mathematical parser, configuration builder, likelihood layer, execution worker, and output system; synthetic parameter-recovery experiments with known fiducial cosmological parameters; and direct comparisons between analyses generated by \textsc{CosmoFit} and independently constructed \textsc{Cobaya} calculations. These tests are designed to verify that the graphical configuration layer does not modify the underlying statistical problem and that the complete workflow reproduces reference posterior constraints.

The principal contributions of the present work are therefore:
\begin{enumerate}
\item A desktop graphical workflow for configuring Bayesian analyses of user-defined cosmological expansion histories;

\item A restricted mathematical-expression parser that avoids the unrestricted execution of user-provided code.

\item Automatic and transparent generation of reproducible \textsc{Cobaya} configurations.

\item Isolated execution of Markov chain Monte Carlo analyses with persistent status and logging information.

\item Integrated posterior inspection and visualization through \textsc{GetDist}.

\item Support for cosmic chronometers and official \textsc{Cobaya} Type Ia supernova likelihoods.

\item Validation through synthetic parameter recovery and comparisons with independent \textsc{Cobaya} calculations.

\end{enumerate}

The source code, documentation, validation material, and reproducible examples
are publicly available through the project repository:
\url{https://github.com/GiovanyCruz/cosmofit}

\section{Cosmological Framework and Model Implementation}
\subsection{Cosmological Background and Expansion History}

CosmoFit is designed to study the background evolution of homogeneous,
isotropic, and spatially flat universes. Within this framework, the spacetime
geometry is described by the Friedmann--Lemaître--Robertson--Walker metric,

\begin{equation}
ds^2
=
-c^2dt^2
+
a^2(t)
\left[
dr^2
+
r^2
\left(
d\theta^2
+
\sin^2\theta\,d\phi^2
\right)
\right].
\label{eq:flrw_metric}
\end{equation}

Here, \(a(t)\) is the scale factor and \(c\) is the speed of light. The
evolution of the scale factor determines the expansion rate of the Universe,
which is characterized by the Hubble parameter,

\begin{equation}
H(t)
=
\frac{\dot{a}(t)}{a(t)},
\label{eq:hubble_parameter}
\end{equation}
where the dot denotes differentiation with respect to cosmic time. The
cosmological redshift is related to the scale factor through

\begin{equation}
1+z
=
\frac{a_0}{a(t)},
\label{eq:redshift_scale_factor}
\end{equation}
where \(a_0\) is the present value of the scale factor. Throughout this work,
the normalization \(a_0=1\) is adopted. In terms of redshift, a general expansion history can be written as

\begin{equation}
H(z)
=
H_0 E(z;\boldsymbol{\theta}),
\label{eq:general_expansion_history}
\end{equation}
where \(H_0\) is the present value of the Hubble parameter,
\(E(z;\boldsymbol{\theta})\) is the dimensionless expansion function, and
\(\boldsymbol{\theta}\) denotes the set of cosmological parameters.
The main theoretical input required by CosmoFit is an explicit mathematical
expression for \(H(z)\). The parameters appearing in this expression may be
defined either as fixed quantities or as sampled parameters. For each sampled
parameter, the user specifies a prior interval, a reference value, and, when
required, a proposal scale through the graphical interface.
This approach is particularly suitable for models whose background dynamics
can be expressed in the form

\begin{equation}
H(z)
=
H(z;\theta_1,\theta_2,\ldots,\theta_N).
\label{eq:user_defined_expansion}
\end{equation}

Therefore, the application can be used with both standard cosmological models
and phenomenological or effective proposals. When a model requires the
simultaneous numerical solution of a system of differential equations, its
solution must first be reduced to an explicit function of redshift before it
can be incorporated into the CosmoFit workflow.

\subsection{User-Defined Expansion Models}

The spatially flat \(\Lambda\)CDM model provides a simple example of how an
expansion history can be introduced into the application. Neglecting the
radiation contribution at late times, the Hubble parameter is given by

\begin{equation}
H_{\Lambda\mathrm{CDM}}(z)
=
H_0
\sqrt{
\Omega_m(1+z)^3
+
1-\Omega_m
},
\label{eq:lcdm_hubble}
\end{equation}
where \(\Omega_m\) is the present matter-density parameter. In CosmoFit, this
expansion history can be entered as

\begin{verbatim}
H0*sqrt(Om*(1+z)**3 + 1-Om)
\end{verbatim}
where \texttt{H0} and \texttt{Om} are declared as model parameters.
CosmoFit is not restricted to the \(\Lambda\)CDM model. The user may
introduce any expression that is compatible with the mathematical syntax
accepted by the application. This flexibility allows the study of effective dark-energy models\cite{amendola2010dark}, modified
gravity theories \cite{de2010f, padmanabhan2013lanczos}, braneworld scenarios \cite{arroyo2026lovelock, dvali20004d}, and other non-standard cosmologies,
provided that their background predictions can be expressed explicitly
through \(H(z)\).

\subsection{Cosmological Observables}

The expansion history can be confronted with different late-time cosmological
observables. In the case of Type Ia supernovae, the comparison with
observational data is performed through cosmological distances. For a
spatially flat universe, the line-of-sight comoving distance is

\begin{equation}
D_C(z)
=
c\int_0^z
\frac{dz'}{H(z')}.
\label{eq:comoving_distance}
\end{equation}

In this case, the transverse comoving distance coincides with \(D_C(z)\), and
the luminosity distance is given by
\begin{equation}
D_L(z)
=
(1+z)D_C(z).
\label{eq:luminosity_distance}
\end{equation}
The corresponding theoretical distance modulus is
\begin{equation}
\mu_{\mathrm{th}}(z)
=
5\log_{10}
\left[
\frac{D_L(z)}{\mathrm{Mpc}}
\right]
+
25.
\label{eq:distance_modulus}
\end{equation}

These relations connect the expansion history defined by the user with the
Type Ia supernova likelihoods supported by the application.

Cosmic chronometers, in contrast, provide direct measurements of the Hubble
parameter at different redshifts. For these data, the intermediate
calculation of cosmological distances is not required. The theoretical
prediction is obtained by directly evaluating

\begin{equation}
H_{\mathrm{th}}(z_i)
=
H(z_i;\boldsymbol{\theta}),
\label{eq:cc_prediction}
\end{equation}
at each observed redshift \(z_i\).
Thus, the same expression for \(H(z)\) can be used both to construct integrated
observables, such as the luminosity distance, and to compare the model
directly with measurements of the expansion rate.

\subsection{Numerical Domain and Scope of the Implementation}

For an expansion history to be physically and numerically consistent,
\(H(z)\) must remain real, finite, and positive throughout the redshift
interval used in the analysis. Parameter combinations that generate singularities, complex values,
undefined operations, or non-finite results cannot be used to evaluate the
observational likelihood correctly. Before starting the sampling process, CosmoFit checks the syntax of the
expression and performs a preliminary numerical evaluation using the
reference parameter values supplied by the user. This procedure allows common
input errors to be identified before a numerical run is created, including
incorrect parentheses, unsupported functions, undefined domains, and
non-finite values. The preliminary evaluation does not guarantee that every region explored
during sampling will be physically valid. Therefore, whenever a parameter
combination produces an inadmissible prediction, the corresponding model
evaluation is rejected by assigning a non-finite likelihood value.

The application
does not compute scalar, vector, or tensor cosmological perturbations, nor
does it use Boltzmann solvers such as \textsc{class} or \textsc{camb}. These restrictions clearly define the scope of the first release. CosmoFit is
not intended to replace general cosmological inference tools operating at the
perturbation level. Its purpose is to provide an accessible and transparent
environment for defining, validating, and confronting background expansion
histories with late-time observational data.

\section{Bayesian inference framework}
\label{sec:bayesian_framework}

\textsc{CosmoFit} connects the expansion history defined by the user with
observational data through a Bayesian analysis. Once the function \(H(z)\),
the model parameters, and the datasets have been specified, the objective is
to determine which parameter values provide the best description of the
observations.

\subsection{Posterior distribution}

Let \(\boldsymbol{\theta}\) denote the set of parameters of a cosmological
model \(\mathcal{M}\), and let \(D\) represent the observational data. In the
Bayesian approach, the information available before the analysis is described
by the prior, while the agreement between the model and the data is quantified
by the likelihood. These two ingredients are combined through Bayes' theorem,

\begin{equation}
    P(\boldsymbol{\theta}\mid D,\mathcal{M})
    =
    \frac{
        \mathcal{L}(D\mid\boldsymbol{\theta},\mathcal{M})
        \pi(\boldsymbol{\theta}\mid\mathcal{M})
    }{
        \mathcal{Z}(D\mid\mathcal{M})
    },
    \label{eq:bayes_theorem}
\end{equation}

where \(\mathcal{L}\) is the likelihood, \(\pi\) is the prior, and
\(\mathcal{Z}\) is the Bayesian evidence. Since the current version of
\textsc{CosmoFit} is intended for parameter estimation within a fixed model,
the evidence acts only as a normalization constant. Therefore,

\begin{equation}
    P(\boldsymbol{\theta}\mid D,\mathcal{M})
    \propto
    \mathcal{L}(D\mid\boldsymbol{\theta},\mathcal{M})
    \pi(\boldsymbol{\theta}\mid\mathcal{M}).
    \label{eq:posterior_proportional}
\end{equation}

The posterior distribution therefore identifies the combinations of
parameters that are compatible with the chosen priors and, at the same time,
provide an adequate description of the observational data.

\subsection{Likelihoods and priors}

The likelihood measures the agreement between the predictions of the model
and the observations. For Gaussian data with covariance matrix
\(\mathbf{C}\), this agreement can be expressed as

\begin{equation}
    \chi^{2}(\boldsymbol{\theta})
    =
    \left[
        \boldsymbol{d}
        -
        \boldsymbol{t}(\boldsymbol{\theta})
    \right]^{\mathrm{T}}
    \mathbf{C}^{-1}
    \left[
        \boldsymbol{d}
        -
        \boldsymbol{t}(\boldsymbol{\theta})
    \right],
    \qquad
    \ln\mathcal{L}
    =
    -\frac{1}{2}\chi^{2}
    +
    \mathrm{constant},
    \label{eq:gaussian_likelihood}
\end{equation}

where \(\boldsymbol{d}\) represents the observational data and
\(\boldsymbol{t}(\boldsymbol{\theta})\) is the corresponding theoretical
prediction. When statistically independent datasets are used, their
contributions to the total log-likelihood are added.

In \textsc{CosmoFit}, the cosmic-chronometer likelihood is evaluated directly
within the application, while the official supernova likelihoods are handled
through their corresponding \textsc{Cobaya} components. The construction of
these likelihoods is described in Sec.~\ref{sec:datasets}.

Each model parameter can be defined as either fixed or sampled. For sampled
parameters, the user specifies a uniform prior, a reference value, and a
proposal scale. The choice of prior ranges is important: an interval that is
too narrow may artificially truncate the posterior, while an excessively
broad interval may reduce sampling efficiency or include regions with no
physical relevance.

\subsection{MCMC sampling}

For most cosmological models, the posterior distribution cannot be obtained
analytically. Markov chain Monte Carlo methods are therefore used to sample it
numerically \cite{Metropolis1953,Hastings1970}.

\textsc{CosmoFit} does not implement its own sampler. The configuration
selected through the interface is converted into a valid \textsc{Cobaya}
input, and the MCMC calculation is performed using its established
implementation \cite{TorradoLewis2021}. In this way, the interface simplifies
the preparation of the analysis without modifying the statistical procedure
used by \textsc{Cobaya}.

For each proposed parameter set \(\boldsymbol{\theta}\), the theoretical
component evaluates

\begin{equation}
    H(z;\boldsymbol{\theta}),
    \label{eq:hubble_sampled_parameters}
\end{equation}

and computes the quantities required by the selected likelihood. Parameter
combinations that produce undefined, complex, or non-finite values are
rejected automatically.

The calculation is executed in a process separated from the graphical
interface. This allows the application to remain responsive during sampling
and enables the user to inspect the run status, review diagnostic messages, or
cancel the execution.

\subsection{Posterior analysis and convergence}

Once the sampling process has finished, the chains are analyzed with
\textsc{GetDist} \cite{Lewis2019GetDist}. From the results panel,
\textsc{CosmoFit} provides marginalized means, standard deviations, credible
intervals, one-dimensional distributions, two-dimensional contours, and
triangle plots. Before these results are calculated, the user may discard an initial fraction
of the stored chain. This removes samples that may still be influenced by the
starting point. The ignored fraction, the selected credible level, and the
plotting options are stored together with the results of the analysis. 

Convergence must be checked before interpreting the posterior constraints. One
of the diagnostics used by \textsc{Cobaya} is the Gelman--Rubin statistic
\cite{GelmanRubin1992}, which compares the variation within the chains and
between them. Convergence is approached as

\begin{equation}
    R-1 \longrightarrow 0.
    \label{eq:gelman_rubin_condition}
\end{equation}

However, this quantity should not be considered in isolation. Since
consecutive samples in an MCMC chain are generally correlated, the number of
effectively independent samples may be smaller than the total chain length.
For this reason, convergence, chain length, and the stability of the
marginalized results should be examined together.

\subsection{Reproducibility}

For each run, \textsc{CosmoFit} preserves the model expression, parameter
definitions, prior intervals, selected likelihoods, the generated
\textsc{Cobaya} input, metadata, execution logs, chains, and posterior
summaries. Two independent MCMC runs are not expected to generate exactly the same
sequence of samples, even when the same configuration is used. In this
context, reproducibility means that the same model, data, priors, and sampler
settings lead to statistically compatible posterior distributions. For this reason, the validation procedure compares posterior means, standard
deviations, credible intervals, convergence information, and the recovery of
fiducial values, rather than requiring the chains to agree sample by sample.

\section{Design and implementation of \textsc{CosmoFit}}
\label{sec:software_design}

\textsc{CosmoFit} was developed as a desktop application for configuring,
running, and inspecting Bayesian analyses of background cosmological models.
Its purpose is to simplify the preparation of a \textsc{Cobaya} analysis
without hiding the physical assumptions, parameter definitions, or numerical
configuration on which the results depend.

The application allows the user to define an explicit expression for \(H(z)\),
declare fixed and sampled parameters, assign prior intervals and reference
values, select an observational likelihood, configure the sampler, and inspect
the resulting chains through \textsc{GetDist}. Each stage is represented
explicitly in the graphical workflow, while the generated configuration files
and numerical outputs remain available for independent inspection.

\subsection{Software architecture}

The software is organized as a set of connected modules rather than as a
single graphical program. Its general workflow can be summarized as

\begin{equation}
\begin{aligned}
    \text{user input}
    &\longrightarrow
    \text{validation}
    \longrightarrow
    \text{\textsc{Cobaya} configuration}
    \\
    &\longrightarrow
    \text{sampling}
    \longrightarrow
    \text{\textsc{GetDist} analysis}.
\end{aligned}
\label{eq:cosmofit_general_workflow}
\end{equation}

The graphical interface is implemented with \textsc{PySide6}, while the
numerical calculations are delegated to established scientific packages,
including \textsc{NumPy}, \textsc{SciPy}, \textsc{Cobaya}, and
\textsc{GetDist}. This separation keeps the interface focused on model
configuration, validation, execution control, and result inspection.

Internally, the information entered through the graphical widgets is converted
into a normalized configuration that is independent of the interface itself.
This representation contains the model expression, parameter definitions,
priors, likelihood selection, sampler options, and output settings. It can be
stored, tested, and translated into the final input required by
\textsc{Cobaya}.

\subsection{Model definition and validation}

The user defines the expansion history by entering an explicit mathematical
expression for \(H(z)\). For example, the spatially flat
\(\Lambda\)CDM model can be written as

\begin{verbatim}
H0*sqrt(Om*(1+z)**3 + 1-Om)
\end{verbatim}
with \texttt{H0} and \texttt{Om} declared in the parameter table. Each
parameter may be fixed at a chosen value or sampled within a specified prior
interval.
The expression is interpreted through a restricted mathematical parser. The
accepted syntax includes the redshift variable \(z\), declared parameter
names, numerical constants, arithmetic operators, parentheses, and a limited
set of mathematical functions. Symbols or operations outside this supported
set are rejected before the analysis is launched. The parser first converts the expression into an abstract syntax tree and then
builds the numerical function used during sampling. This makes it possible to
check the structure of the model before passing it to the inference engine. The application also evaluates the model at the supplied reference values over
a selected redshift interval. This preliminary test can reveal divisions by
zero, invalid square roots or logarithms, overflows, complex results, and
non-finite values. Although this check cannot guarantee that every point
visited by the sampler will be valid, it detects many common configuration
errors before numerical execution begins. Invalid parameter combinations
encountered during sampling are assigned a non-finite likelihood and rejected.

\subsection{Parameter configuration and \textsc{Cobaya} input}

Each parameter is stored in a normalized form. A fixed parameter keeps the
same value throughout the run, whereas a sampled parameter is assigned a
uniform prior,

\begin{equation}
    \theta_i^{\mathrm{min}}
    \leq
    \theta_i
    \leq
    \theta_i^{\mathrm{max}},
    \label{eq:cosmofit_parameter_interval}
\end{equation}
together with a reference value and a proposal scale. Before the configuration is accepted, \textsc{CosmoFit} verifies that the
prior bounds are properly ordered and that the reference value lies inside the
selected interval. The validated definitions are then translated into the
parameter block used by \textsc{Cobaya}.
The complete translation follows the sequence

\begin{equation}
    \text{graphical state}
    \longrightarrow
    \text{normalized configuration}
    \longrightarrow
    \text{\textsc{Cobaya} input}.
    \label{eq:configuration_translation}
\end{equation}

The generated input includes the fixed and sampled parameters, prior bounds,
reference values, proposal scales, likelihood components, external theoretical
function, sampler settings, output prefix, and package paths required by
external datasets. Since this file is stored with the run, the calculation can
be inspected or repeated independently of the graphical application.

\subsection{Likelihood integration}

The current implementation supports an internal cosmic-chronometer likelihood
and the official Pantheon+, Pantheon+SH0ES, and Union3 supernova likelihoods
available through \textsc{Cobaya}.
\\For cosmic chronometers, \textsc{CosmoFit} reads a table with the columns

\begin{verbatim}
z,H,sigma
\end{verbatim}
and evaluates the user-defined model at the observed redshifts. The data loader
checks the column names, array dimensions, finite entries, and positive
uncertainties before the likelihood is constructed.

For the supernova compilations, the corresponding likelihood calculations are
delegated to the official \textsc{Cobaya} components. Their external data
packages are installed separately and are not redistributed with
\textsc{CosmoFit}. This avoids duplicating likelihood implementations that are
already maintained within the \textsc{Cobaya} ecosystem. The observational
datasets and their statistical treatment are discussed in
Sec.~\ref{sec:datasets}.

\subsection{Execution and posterior analysis}

MCMC calculations may require many evaluations of the theoretical model and
can remain active for an extended period. To prevent the graphical interface
from becoming unresponsive, the analysis is executed in a worker process
separate from the main application.
The worker receives the generated configuration, runs \textsc{Cobaya}, writes
the chains and diagnostic messages, and records whether the calculation
finished successfully. This separation also makes it possible to cancel an
active run and to examine the reason for a failure after the interface has
been closed.
Once the sampling process is complete, the chains can be loaded and analyzed
with \textsc{GetDist}. The user may calculate marginalized means, standard
deviations, and credible intervals, as well as generate one-dimensional,
two-dimensional, and triangle plots. Because posterior processing is separated
from sampling, the same chains can be analyzed repeatedly with different
ignored fractions, confidence levels, parameter selections, or plotting
options without running the MCMC calculation again.
The numerical summaries are stored in machine-readable form, so the graphical
plots are not the only record of the posterior results. They can therefore be
used in independent comparisons, tables, or validation scripts.

\subsection{Run outputs and reproducibility}

Each analysis is stored in a dedicated directory containing the scientific
configuration and the products generated during execution. A typical run
includes

\begin{verbatim}
input.yaml
normalized_config.json
cobaya_input.yaml
metadata.json
status.json
summary.json
logs/
chains/
\end{verbatim}

The file \texttt{input.yaml} records the scientific choices entered by the
user, while \texttt{normalized\_config.json} contains their validated internal
representation. The file \texttt{cobaya\_input.yaml} stores the configuration
passed to \textsc{Cobaya}. Runtime information and execution state are written
to \texttt{metadata.json} and \texttt{status.json}, whereas
\texttt{summary.json} contains the posterior statistics generated after
processing the chains. The directories \texttt{logs/} and \texttt{chains/}
store diagnostic messages and MCMC samples, respectively.
This structure allows a completed or failed analysis to be examined without
depending on the temporary state of the interface. The model expression,
parameter ranges, selected likelihood, sampler configuration, execution
status, and posterior products remain associated with the same run.
Reproducibility in an MCMC calculation does not require two independent runs
to generate identical chains. Rather, analyses performed with the same model,
data, priors, and sampler settings should lead to statistically compatible
posterior distributions. Fixed random seeds are used in the reference
validation workflows whenever they are supported by the underlying
calculation.

\subsection{Distribution and requirements}

The current version of \textsc{CosmoFit} is distributed as source code through
its public repository. Running the application requires \textsc{Git},
Python~3.12, and the project dependencies, including \textsc{PySide6},
\textsc{Cobaya}, \textsc{GetDist}, \textsc{NumPy}, \textsc{SciPy}, and
\textsc{PyYAML}. Testing and development additionally use tools such as
\textsc{pytest} and \textsc{Ruff}.
\\A standard installation can be performed with

\begin{verbatim}
git clone https://github.com/GiovanyCruz/cosmofit.git
cd cosmofit
python3.12 -m venv .venv
source .venv/bin/activate
python -m pip install --upgrade pip
python -m pip install -e .
\end{verbatim}
After installation, the interface can be launched with

\begin{verbatim}
cosmofit
\end{verbatim}
or with

\begin{verbatim}
python -m cosmofit.ui
\end{verbatim}
The external data packages required by the official supernova likelihoods must
be installed through \textsc{Cobaya}. Additional installation, dataset, and
troubleshooting instructions are provided in the project documentation.
\section{Observational datasets and likelihoods}
\label{sec:datasets}

The current version of \textsc{CosmoFit} includes observational probes of the
late-time expansion of the Universe. It supports direct measurements of the
Hubble parameter from cosmic chronometers and the official \textsc{Cobaya}
likelihoods for the Pantheon+, Pantheon+SH0ES, and Union3 Type Ia supernova
compilations. These probes are complementary. Cosmic chronometers constrain the expansion
rate \(H(z)\) directly, whereas Type Ia supernovae constrain the
luminosity-distance--redshift relation. Both can therefore be connected to a
user-defined background expansion history without requiring the calculation
of cosmological perturbations.

\subsection{Cosmic chronometers}
\label{subsec:cosmic_chronometers}

The cosmic-chronometer method estimates the Hubble parameter from the
differential age evolution of passively evolving galaxies
\cite{JimenezLoeb2002,Moresco2012,moresco2016constraining,moresco2020setting}.
It is based on the relation

\begin{equation}
    H(z)
    =
    -\frac{1}{1+z}
    \frac{dz}{dt},
    \label{eq:cosmic_chronometer_relation}
\end{equation}
which allows \(H(z)\) to be inferred without first integrating the expansion
history to obtain a cosmological distance.
For a set of measurements
\(\{z_i,H_{\mathrm{obs}}(z_i),\sigma_{H,i}\}\), \textsc{CosmoFit} evaluates
the user-defined model at the observed redshifts and computes

\begin{equation}
    \chi_{\mathrm{CC}}^{2}(\boldsymbol{\theta})
    =
    \sum_{i}
    \frac{
        \left[
            H_{\mathrm{obs}}(z_i)
            -
            H(z_i;\boldsymbol{\theta})
        \right]^{2}
    }{
        \sigma_{H,i}^{2}
    }.
    \label{eq:cc_chi_square_diagonal}
\end{equation}

This expression corresponds to statistically independent measurements. The
normalization of the Gaussian likelihood is constant when the uncertainties do
not depend on the cosmological parameters and therefore does not affect the
relative posterior probabilities.
The graphical workflow accepts tabular files with the columns

\begin{verbatim}
z,H,sigma
\end{verbatim}
where \(H\) and \(\sigma\) are expressed in
\(\mathrm{km\,s^{-1}\,Mpc^{-1}}\). Before the likelihood is created, the
loader checks that the required columns are present, the arrays have consistent
lengths, all entries are finite, and the uncertainties are positive. This internal likelihood is also used in the synthetic-recovery and
independent-comparison tests described in Sec.~\ref{sec:validation}.

\subsection{Type Ia supernovae}
\label{subsec:type_ia_supernovae}

Type Ia supernovae are standardizable candles that constrain the relation
between luminosity distance and redshift. For the spatially flat cosmologies
considered in the present release, the theoretical luminosity distance is

\begin{equation}
    D_{L}(z;\boldsymbol{\theta})
    =
    c(1+z)
    \int_{0}^{z}
    \frac{dz'}{H(z';\boldsymbol{\theta})}.
    \label{eq:sn_luminosity_distance}
\end{equation}

The corresponding distance modulus follows from

\begin{equation}
    \mu_{\mathrm{th}}(z;\boldsymbol{\theta})
    =
    5\log_{10}
    \left[
        \frac{D_L(z;\boldsymbol{\theta})}{\mathrm{Mpc}}
    \right]
    +25.
    \label{eq:sn_theoretical_distance_modulus}
\end{equation}

The precise construction of the supernova likelihood depends on the selected
compilation, including its covariance matrix, calibration information, and
treatment of nuisance parameters. Rather than reproducing these procedures,
\textsc{CosmoFit} provides the background predictions required by the
official likelihood components distributed through \textsc{Cobaya}
\cite{TorradoLewis2021}. This choice avoids maintaining independent copies of complex observational
likelihoods and makes it possible to compare a run configured through
\textsc{CosmoFit} with a direct \textsc{Cobaya} calculation using the same
component.

\subsubsection{Pantheon+}

Pantheon+ contains 1701 light curves corresponding to 1550 distinct Type Ia
supernovae over an approximate redshift range
\(0.001<z<2.26\). The compilation incorporates improved photometric
calibration, redshift information, peculiar-velocity corrections,
selection-bias corrections, and systematic covariance modelling
\cite{scolnic2022pantheon+,brout2022pantheon+}. It is accessed through the official component

\begin{verbatim}
sn.pantheonplus
\end{verbatim}
In the absence of an external absolute calibration, this likelihood mainly
constrains relative distances and the shape of the expansion history. The
overall normalization remains degenerate with the supernova absolute
magnitude, so a Pantheon+-only analysis should not normally be interpreted as
an independent measurement of \(H_{0}\).

\subsubsection{Pantheon+SH0ES}

Pantheon+SH0ES combines the Pantheon+ sample with Cepheid-host calibration
information from the SH0ES distance ladder
\cite{brout2022pantheon+,riess2022comprehensive}. This additional information
breaks the relative-distance normalization degeneracy and allows the absolute
expansion scale to be constrained within the assumptions of the likelihood. The corresponding component is

\begin{verbatim}
sn.pantheonplusshoes
\end{verbatim}
\textsc{CosmoFit} preserves the distinction between Pantheon+ and
Pantheon+SH0ES because the two likelihoods do not contain the same
information. The calibration and nuisance-parameter treatment are those
implemented by the official \textsc{Cobaya} component.

\subsubsection{Union3}

Union3 contains approximately 2000 cosmologically useful Type Ia supernovae
analyzed with the UNITY hierarchical Bayesian framework
\cite{Rubin2025Union3}. It combines observations from several surveys using a
common treatment of calibration, selection effects, population distributions,
and systematic uncertainties. Within \textsc{CosmoFit}, the likelihood is selected through

\begin{verbatim}
sn.union3
\end{verbatim}
As in the Pantheon+ cases, \textsc{CosmoFit} supplies the expansion and
distance predictions, while the observational and statistical treatment
remains that of the official likelihood.

\subsection{External packages and model compatibility}
\label{subsec:external_likelihood_packages}

The supernova datasets are not redistributed with \textsc{CosmoFit}. Their
official data packages must be installed through \textsc{Cobaya}, for example,

\begin{verbatim}
cobaya-install \
  sn.pantheonplus \
  sn.pantheonplusshoes \
  sn.union3 \
  --packages-path "$COBAYA_PACKAGES_PATH"
\end{verbatim}
The selected likelihood and the package path are recorded in the generated run
configuration, while the observational files remain subject to their original
distribution, citation, and licensing conditions. The theoretical quantity required by the likelihood depends on the selected
probe. Cosmic chronometers use \(H(z;\boldsymbol{\theta})\) directly, whereas
supernova likelihoods require the luminosity distance obtained by integrating
\(1/H(z)\). Consequently, a model is compatible with the present likelihoods
only if it produces a real, positive, and finite expansion rate over the full
redshift range of the selected dataset. The preliminary model preview checks the expression at a finite set of
redshifts and reference parameter values. During sampling, the likelihood
performs the required evaluations over the actual dataset range. Parameter
combinations that produce invalid predictions are assigned a non-finite
log-likelihood and rejected.

\subsection{Current observational scope}
\label{subsec:current_observational_scope}

The present release is restricted to background observables. It does not yet
include baryon acoustic oscillations, cosmic microwave background likelihoods,
redshift-space distortions, weak lensing, structure-growth measurements, or
perturbation-level predictions from \textsc{CLASS} or \textsc{CAMB}. This scope is consistent with the current design of the application, whose
main theoretical input is an explicit expression for \(H(z)\). Additional
background probes and perturbation-level interfaces may be incorporated in
future versions.
\section{Verification and numerical validation}
\label{sec:validation}

The reliability of \textsc{CosmoFit} was assessed at both the software and
scientific levels. The validation program combines automated tests,
synthetic-data recovery, and comparisons with independently configured
\textsc{Cobaya} analyses. Together, these checks cover the complete workflow,

\begin{equation}
\begin{aligned}
    \text{model definition}
    &\longrightarrow
    \text{configuration}
    \longrightarrow
    \text{likelihood}
    \\
    &\longrightarrow
    \text{sampling}
    \longrightarrow
    \text{posterior analysis}.
\end{aligned}
\label{eq:validation_complete_workflow}
\end{equation}

The purpose of these tests is to verify that the graphical layer correctly
translates the user configuration into the numerical problem executed by
\textsc{Cobaya}, and that the resulting posterior distributions are
consistent with independent calculations.

\subsection{Automated tests and end-to-end execution}
\label{subsec:automated_tests}

The repository includes unit and integration tests executed with
\textsc{pytest}. They cover the expression parser, parameter validation,
normalized configurations, generation of \textsc{Cobaya} inputs, loading of
cosmic-chronometer data, likelihood evaluation, communication with the
theoretical provider, isolated execution, run artifacts, and posterior
processing with \textsc{GetDist}.
\\The complete test suite can be run with

\begin{verbatim}
pytest -q
\end{verbatim}

One of the integration tests compares the cosmic-chronometer likelihood
evaluated through the provider mechanism with a direct evaluation of the same
model and dataset. Agreement between both values verifies that the theoretical
prediction is passed correctly to the likelihood component.
\\The repository also includes a smoke test that executes a minimal flat
\(\Lambda\)CDM analysis from configuration generation to posterior output:

\begin{verbatim}
python -m cosmofit.cobaya_engine.smoke_run
\end{verbatim}
A successful execution creates the normalized configuration, the generated
\textsc{Cobaya} input, metadata, logs, chains, status information, and the
posterior summary. The reference smoke run completed with

\begin{equation}
    \texttt{exit\_code}=0,
    \label{eq:smoke_test_exit_code}
\end{equation}

confirming that the principal components can operate together in a complete
analysis.

\subsection{Reference model and synthetic recovery}
\label{subsec:synthetic_parameter_recovery}

The principal validation model is the spatially flat \(\Lambda\)CDM
background,

\begin{equation}
    H(z)
    =
    H_{0}
    \sqrt{
        \Omega_{m}(1+z)^{3}
        +
        1-\Omega_{m}
    }.
    \label{eq:validation_lcdm_model}
\end{equation}
Within the interface, it is entered as

\begin{verbatim}
H0*sqrt(Om*(1+z)**3 + 1-Om)
\end{verbatim}
A synthetic cosmic-chronometer dataset containing 15 redshift points was
generated with the fiducial parameters

\begin{equation}
    H_{0}^{\mathrm{fid}}
    =
    70\,
    \mathrm{km\,s^{-1}\,Mpc^{-1}},
    \qquad
    \Omega_{m}^{\mathrm{fid}}
    =
    0.30.
    \label{eq:synthetic_fiducial_parameters}
\end{equation}

The dataset is stored in

\begin{verbatim}
validation/data/
cosmic_chronometers_synthetic_recovery.csv
\end{verbatim}
This experiment tests the complete sequence of data loading, model evaluation,
likelihood calculation, MCMC sampling, chain processing, and posterior
summarization.
\\For the Hubble constant, the recovered marginalized constraint was

\begin{equation}
    H_{0}
    =
    71.061649
    \pm
    1.582522\,
    \mathrm{km\,s^{-1}\,Mpc^{-1}}.
    \label{eq:synthetic_h0_result}
\end{equation}

The \(68\%\) and \(95\%\) credible intervals were, respectively,

\begin{equation}
\begin{aligned}
    69.526532
    &<
    H_{0}
    <
    72.640599,
    \\
    67.858860
    &<
    H_{0}
    <
    74.096396,
\end{aligned}
\qquad
\mathrm{km\,s^{-1}\,Mpc^{-1}}.
\label{eq:synthetic_h0_intervals}
\end{equation}
The fiducial value \(H_{0}^{\mathrm{fid}}=70\,
\mathrm{km\,s^{-1}\,Mpc^{-1}}\) lies within both intervals. Its difference
from the posterior mean is approximately \(0.67\) posterior standard
deviations. The fiducial value \(\Omega_m^{\mathrm{fid}}=0.30\) was likewise
recovered within the \(68\%\) and \(95\%\) marginalized credible regions.
These results show that the complete graphical workflow can recover the known
parameters of the synthetic reference cosmology.

\subsection{Comparison with an independent \textsc{Cobaya} analysis}
\label{subsec:independent_cobaya_comparison}

A more direct validation was carried out by comparing two independently
constructed analyses:

\begin{enumerate}
    \item a run generated and executed through \textsc{CosmoFit};
    \item a direct \textsc{Cobaya} run created without using the
    \textsc{CosmoFit} configuration builder.
\end{enumerate}

Both calculations used the same flat \(\Lambda\)CDM model, cosmic-chronometer
dataset, parameter definitions, priors, and posterior-processing prescription.
The first \(30\%\) of each stored chain was discarded before calculating the
marginalized statistics.

The resulting constraints are shown in
Table~\ref{tab:cosmofit_direct_cobaya_comparison}.

\begin{table*}[htbp]
    \centering
    \caption{
        Comparison between the posterior constraints obtained with
        \textsc{CosmoFit} and an independently configured direct
        \textsc{Cobaya} run.
    }
    \label{tab:cosmofit_direct_cobaya_comparison}
    \begin{tabular}{lcccc}
        \hline
        Parameter &
        \textsc{CosmoFit} mean &
        \textsc{CosmoFit} \(\sigma\) &
        Direct \textsc{Cobaya} mean &
        Direct \textsc{Cobaya} \(\sigma\)
        \\
        \hline
        \(H_{0}\,[\mathrm{km\,s^{-1}\,Mpc^{-1}}]\)
        & \(72.344391\)
        & \(1.102487\)
        & \(72.346278\)
        & \(1.036889\)
        \\
        \(\Omega_{m}\)
        & \(0.234969\)
        & \(0.015398\)
        & \(0.234712\)
        & \(0.014701\)
        \\
        \hline
    \end{tabular}
\end{table*}
The relative differences between the posterior means were

\begin{equation}
    \delta_{\mu}(H_{0})=0.0026\%,
    \qquad
    \delta_{\mu}(\Omega_{m})=0.1098\%,
    \label{eq:validation_relative_mean_results}
\end{equation}
while the differences between the posterior standard deviations were

\begin{equation}
    \delta_{\sigma}(H_{0})=6.1324\%,
    \qquad
    \delta_{\sigma}(\Omega_{m})=4.6342\%.
    \label{eq:validation_relative_sigma_results}
\end{equation}
The posterior means agree well below the one-percent level. The smaller
differences in posterior width are compatible with the expected variation
between finite MCMC realizations, including the effects of autocorrelation,
burn-in selection, and numerical estimation of marginalized distributions.
\\The comparison can be reproduced with

\begin{verbatim}
validation/compare_cc_runs.py
\end{verbatim}

and its numerical output is stored in

\begin{verbatim}
validation/results/cc_lcdm_comparison.txt
\end{verbatim}

\subsection{Official supernova likelihoods}
\label{subsec:supernova_validation}

The validation procedure was also applied to the official Pantheon+,
Pantheon+SH0ES, and Union3 likelihood components. For each case, a
\textsc{CosmoFit}-generated analysis was compared with an independently
written \textsc{Cobaya} configuration using the same model, priors,
likelihood, and posterior-processing choices.
\\The consolidated results and the complete protocol are stored in

\begin{verbatim}
validation/results/validation_summary.csv
validation/results/validation_summary.md
docs/validation_protocol.md
\end{verbatim}
This extends the validation beyond the internal cosmic-chronometer likelihood
and checks that the background quantities produced by \textsc{CosmoFit} can
be passed consistently to the official external likelihood components.

\subsection{Posterior processing and reproducibility}
\label{subsec:validation_reproducibility}

The same \textsc{GetDist} settings were used when comparing the
\textsc{CosmoFit} and direct \textsc{Cobaya} chains. In the reference
cosmic-chronometer analysis, an initial fraction

\begin{equation}
    f_{\mathrm{burn}}=0.30
    \label{eq:validation_burn_fraction}
\end{equation}
was excluded from both chains. A small diagnostic outlier fraction of
approximately \(1.4\times10^{-3}\) was found during processing and had no
appreciable effect on the central posterior constraints.
The validation runs preserve the model expression, parameter definitions,
priors, likelihood selection, sampler configuration, random seed when
applicable, generated inputs, chains, logs, and numerical summaries. Fixed
seeds are used whenever supported, although independent MCMC runs are not
expected to produce identical chains.
Reproducibility is therefore assessed through statistical agreement: analyses
performed with the same scientific and numerical configuration should recover
compatible posterior means, widths, credible intervals, and fiducial
parameters.
Overall, the validation program shows that the principal software components
operate as intended, that the complete workflow recovers the parameters of a
known synthetic cosmology, and that analyses configured through
\textsc{CosmoFit} produce posterior constraints consistent with independent
\textsc{Cobaya} calculations. These tests do not guarantee that every
user-defined model is physically meaningful or numerically stable, but they
show that the interface preserves the underlying inference problem for the
models and likelihoods considered in this work.

\section{Illustrative \texorpdfstring{\(w\)CDM}{wCDM} workflow}
\label{sec:illustrative_workflow}

To illustrate the complete workflow of \textsc{CosmoFit}, we consider a
spatially flat \(w\)CDM model constrained with the calibrated
Pantheon+SH0ES supernova likelihood. The aim of this example is to show how a
user-defined expansion history is entered, validated, sampled, and analyzed
within the same graphical environment.

Neglecting radiation at late times, the expansion history is

\begin{equation}
    H(z)
    =
    H_{0}
    \left[
        \Omega_{m}(1+z)^{3}
        +
        \left(1-\Omega_{m}\right)(1+z)^{3(1+w)}
    \right]^{1/2},
    \label{eq:illustrative_wcdm_hubble}
\end{equation}
where \(H_{0}\) is the present Hubble parameter, \(\Omega_{m}\) is the
present matter-density parameter, and \(w\) is the constant dark-energy
equation-of-state parameter. The \(\Lambda\)CDM limit is recovered for
\(w=-1\).

\subsection{Model and parameter configuration}
\label{subsec:wcdm_model_and_parameters}
The expansion history is entered directly in the model panel as

\begin{figure*}[htbp]
    \centering
    \includegraphics[width=0.7\textwidth]{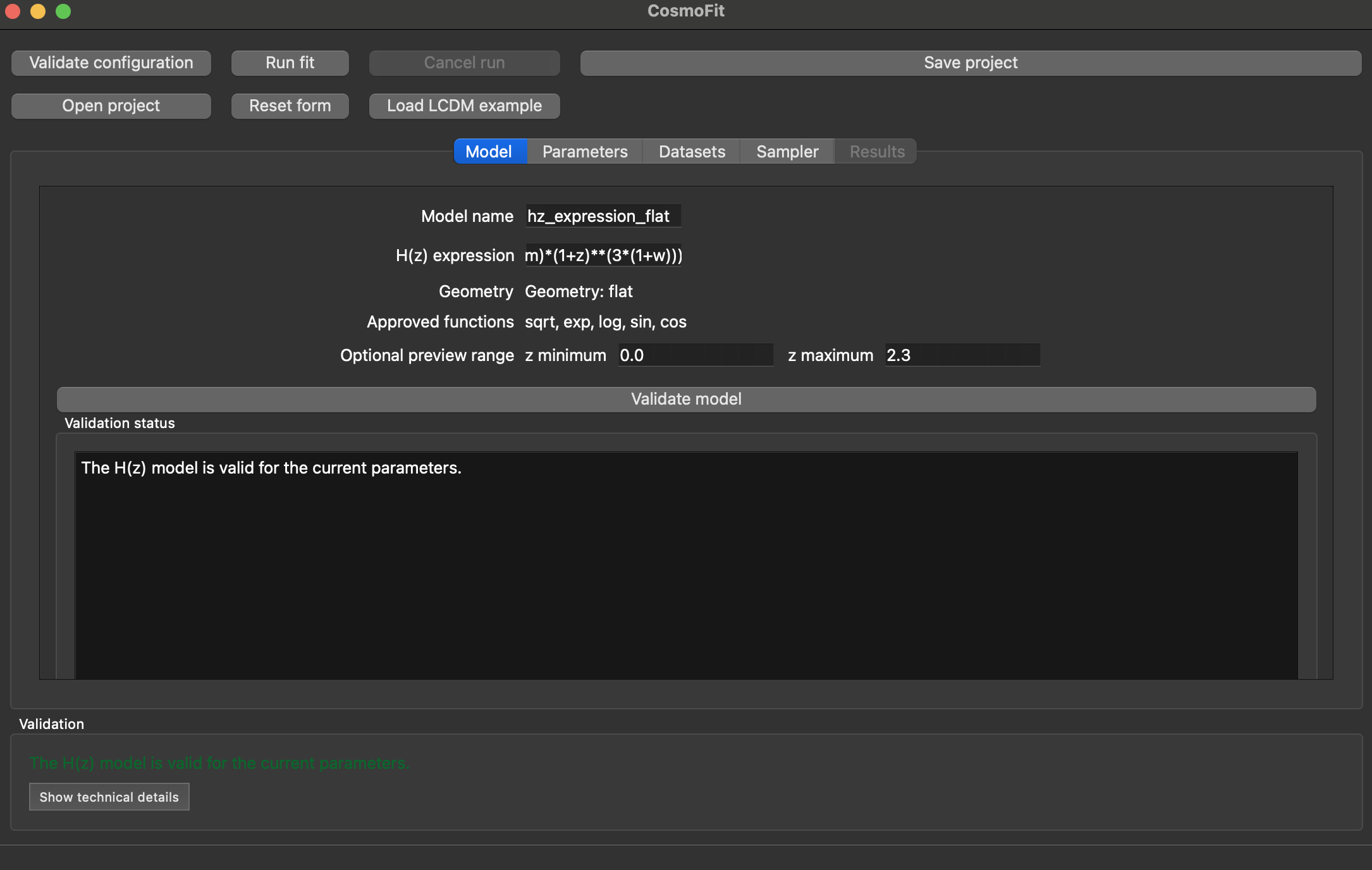}
    \caption{
        Definition and validation of the spatially flat \(w\)CDM expansion
        history. The model is evaluated over an optional redshift interval
        before the remaining analysis settings are configured.
    }
    \label{fig:wcdm_model_definition}
\end{figure*}

\begin{verbatim}
H0*sqrt(Om*(1+z)**3 + (1-Om)*(1+z)**(3*(1+w)))
\end{verbatim}
The symbols \texttt{H0}, \texttt{Om}, and \texttt{w} are then declared in
the parameter table. The model was previewed over the interval
\(0\leq z\leq2.3\), which approximately covers the redshift range relevant
to the selected supernova sample. This interval is used only for numerical
validation and does not impose a cut on the likelihood data. The validation step checks that the expression uses declared parameters and
supported mathematical functions, and that it produces real and finite values
for the reference configuration. Figure~\ref{fig:wcdm_model_definition}
shows the successfully validated model.

\begin{figure*}[htbp]
    \centering
    \includegraphics[width=0.7\textwidth]{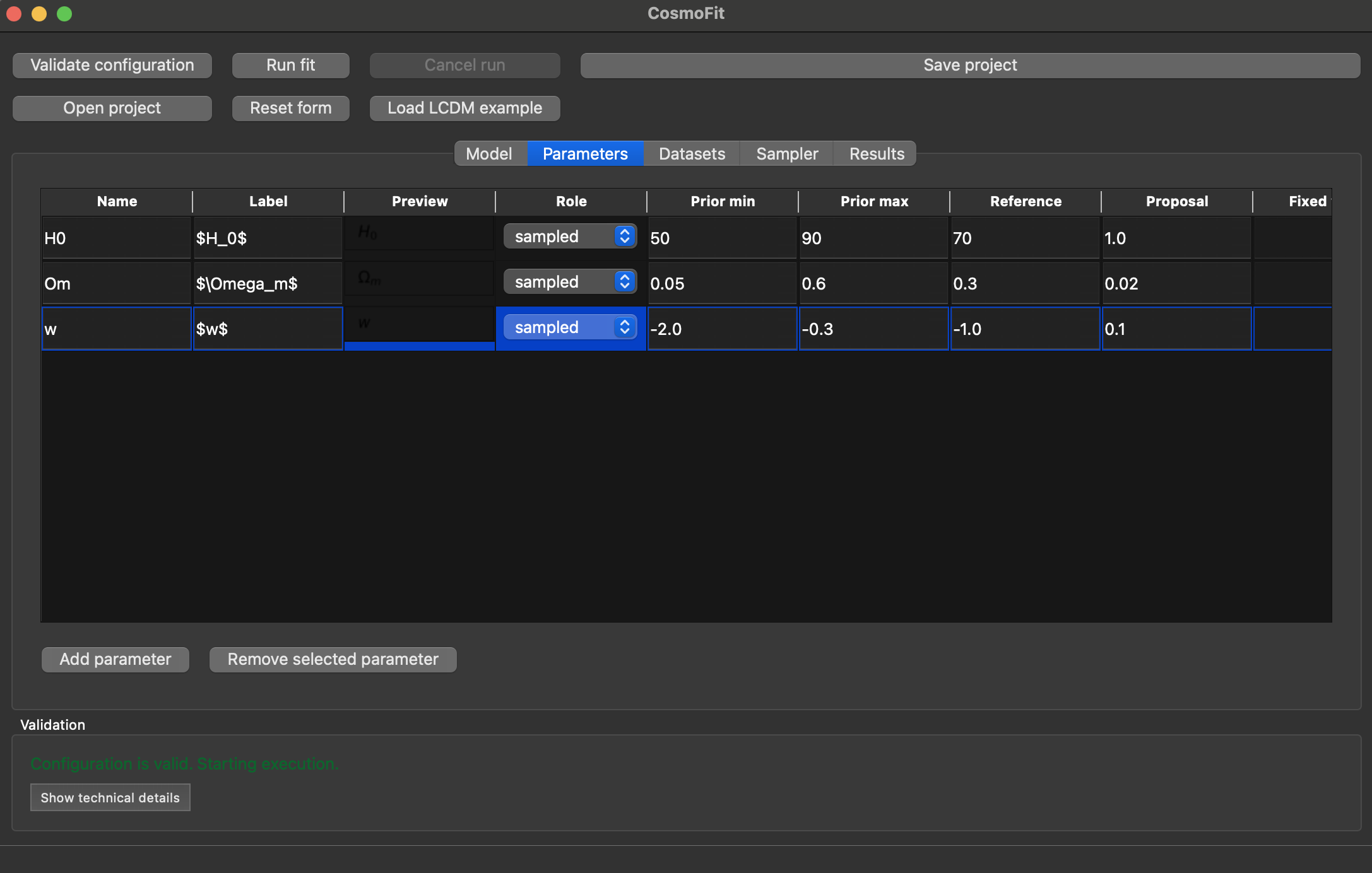}
    \caption{
        Parameter and prior configuration for the illustrative analysis. The
        parameters \(H_{0}\), \(\Omega_{m}\), and \(w\) are sampled with
        uniform priors, reference values, and initial proposal widths.
    }
    \label{fig:wcdm_parameter_configuration}
\end{figure*}

\FloatBarrier
All three parameters were treated as sampled quantities. Their prior ranges,
reference values, and initial proposal widths are  displayed graphically in
Fig.~\ref{fig:wcdm_parameter_configuration}.

\FloatBarrier
The explicit parameter names must match those appearing in the expression for
\(H(z)\), while the mathematical labels are used only in plots and numerical
summaries.

\subsection{Likelihood and sampler settings}
\label{subsec:wcdm_likelihood_and_sampler}
\begin{figure*}[htbp]
    \centering
    \includegraphics[width=0.8\textwidth]{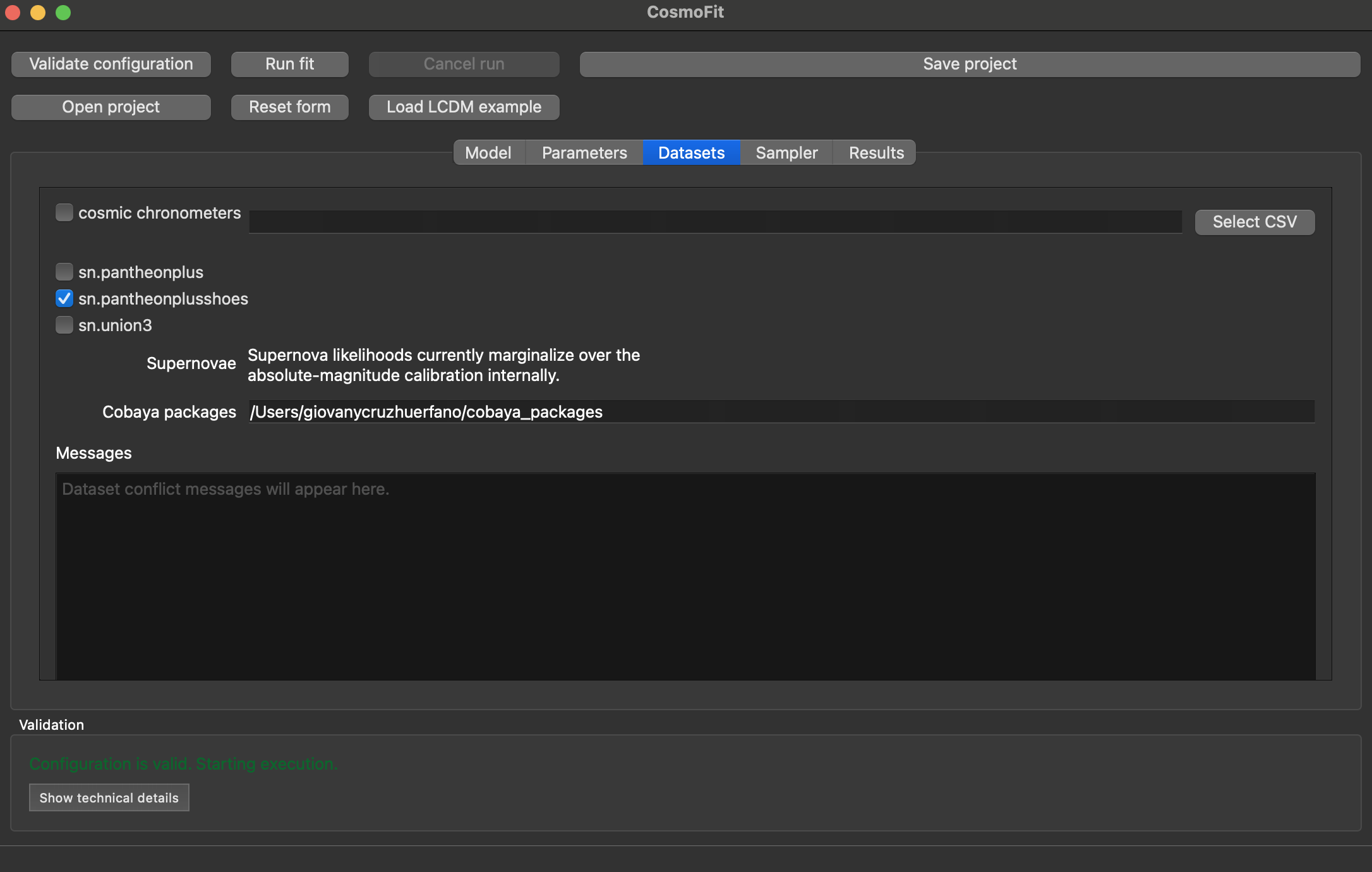}
    \caption{
        Selection of the Pantheon+SH0ES likelihood for the illustrative
        \(w\)CDM run. The panel also displays the local
        \textsc{Cobaya} packages directory used to locate the external data.
    }
    \label{fig:wcdm_dataset_selection}
\end{figure*}

The calibrated Pantheon+SH0ES likelihood was selected through the official
\textsc{Cobaya} component

\begin{verbatim}
sn.pantheonplusshoes
\end{verbatim}

as shown in Fig.~\ref{fig:wcdm_dataset_selection}. The corresponding external
data are located through the local \textsc{Cobaya} packages directory.
\textsc{CosmoFit} supplies the expansion history and luminosity-distance
predictions, while the calibration and nuisance-parameter treatment remain
those of the official likelihood.

\FloatBarrier
The sampling calculation was performed with the MCMC implementation of
\textsc{Cobaya}. The main settings were

\begin{equation}
    s=1234,
    \qquad
    N_{\mathrm{max}}=10\,000,
    \qquad
    R-1<0.01,
    \qquad
    R_{\mathrm{cl}}-1<0.2.
    \label{eq:wcdm_sampler_settings}
\end{equation}

\begin{figure*}[htbp]
    \centering
    \includegraphics[width=0.8\textwidth]{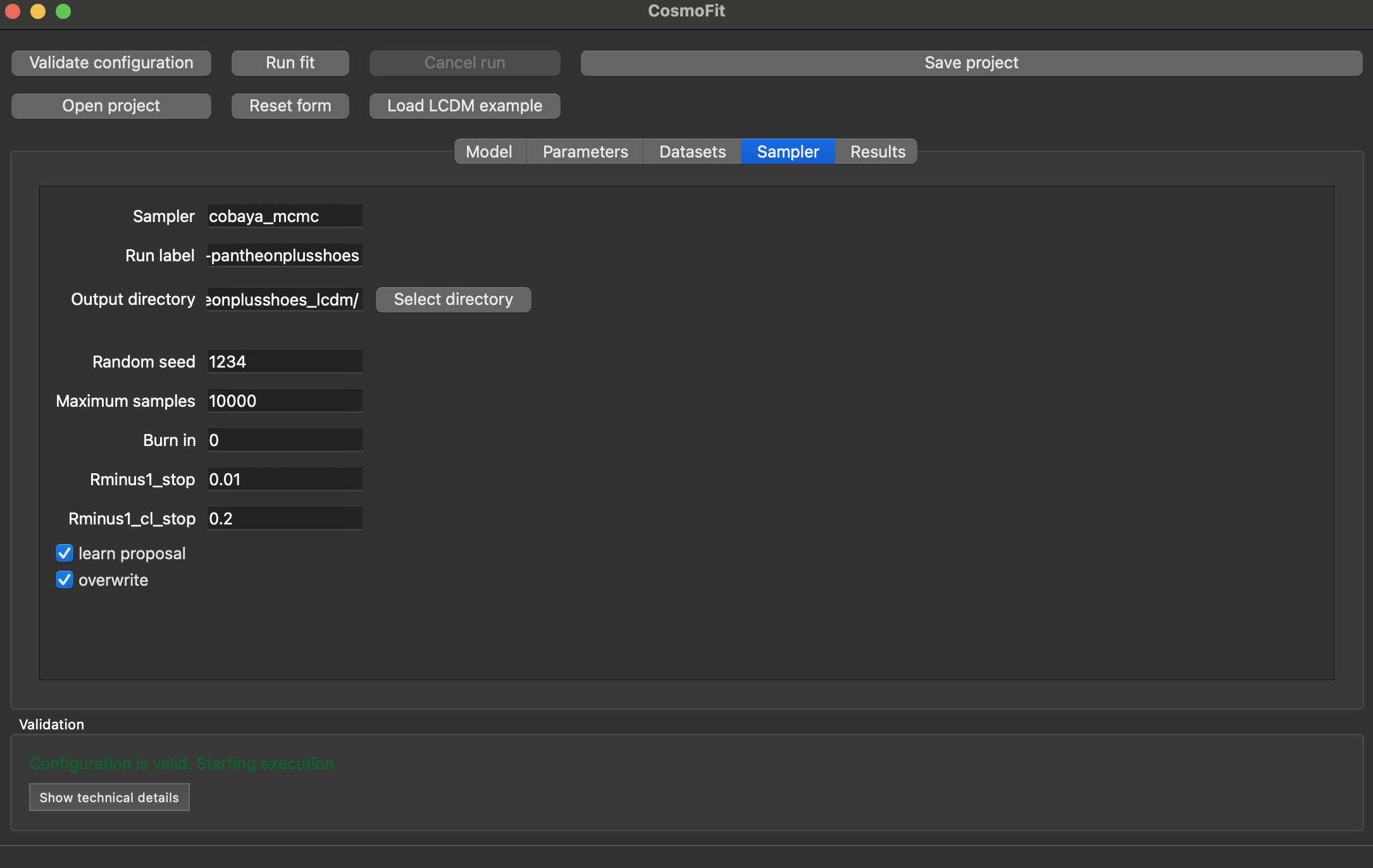}
    \caption{
        Sampler configuration for the \(w\)CDM analysis. The panel records
        the random seed, maximum number of samples, convergence thresholds,
        output location, and proposal-learning option.
    }
    \label{fig:wcdm_sampler_configuration}
\end{figure*}

Proposal learning was enabled so that the sampler could adapt to correlations
among the parameters. The sampler-level burn-in was set to zero because the
initial portion of the chain was removed later during posterior processing.
The complete configuration is shown in
Fig.~\ref{fig:wcdm_sampler_configuration}.

\FloatBarrier
\subsection{Execution of the analysis}
\label{subsec:wcdm_run_execution}

After validating the complete configuration, the analysis is launched with
the \texttt{Run fit} action. \textsc{CosmoFit} then creates the run directory,
writes the generated \textsc{Cobaya} input, and starts the MCMC calculation
in a worker process separated from the graphical interface.

During execution, the results panel reports the state as
\texttt{Running}, displays the output directory, and shows diagnostic
messages produced by the background-theory provider and the
\textsc{Cobaya} worker. Posterior-processing actions remain disabled until
the calculation has finished.

\begin{figure*}[htbp]
    \centering
    \includegraphics[width=0.8\textwidth]{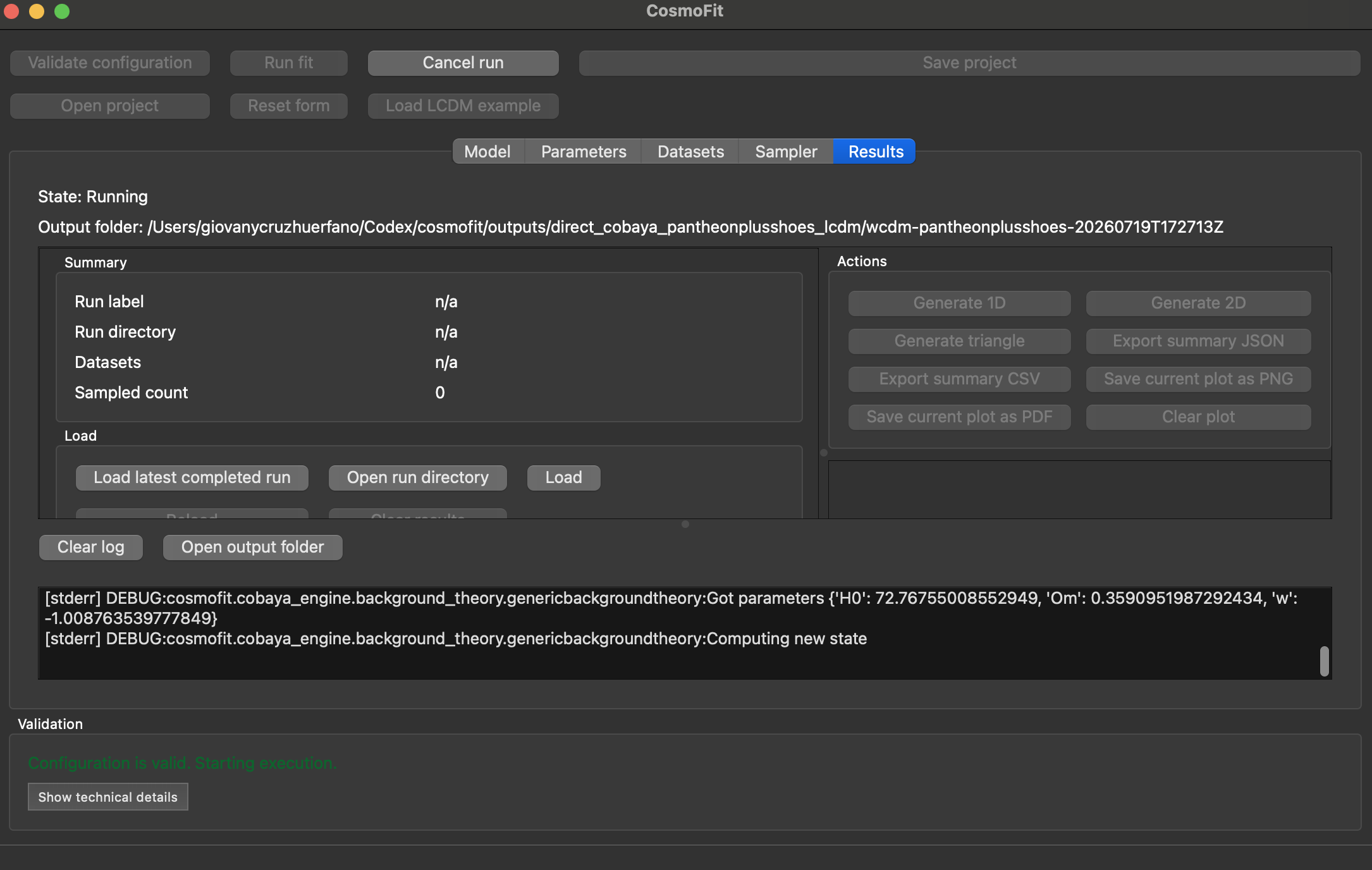}
    \caption{
        Execution of the illustrative \(w\)CDM analysis after pressing
        \texttt{Run fit}. The interface reports the active run state,
        output directory, and diagnostic messages while posterior actions
        remain disabled.
    }
    \label{fig:wcdm_run_execution}
\end{figure*}
\FloatBarrier
The separate worker process keeps the interface responsive and allows the
calculation to be cancelled without closing the application. The run finishes
when the convergence conditions are satisfied or the maximum sample limit is
reached.

\subsection{Posterior processing}
\label{subsec:wcdm_posterior_processing}

After successful completion, the run state changes to \texttt{Ready}. The
reference execution ended after 2280 accepted MCMC steps. The chains were then
processed with \textsc{GetDist}, discarding the first \(30\%\) of the stored
samples and using a primary credible level of \(68\%\).

\begin{figure*}[htbp]
    \centering
    \includegraphics[width=0.8\textwidth]{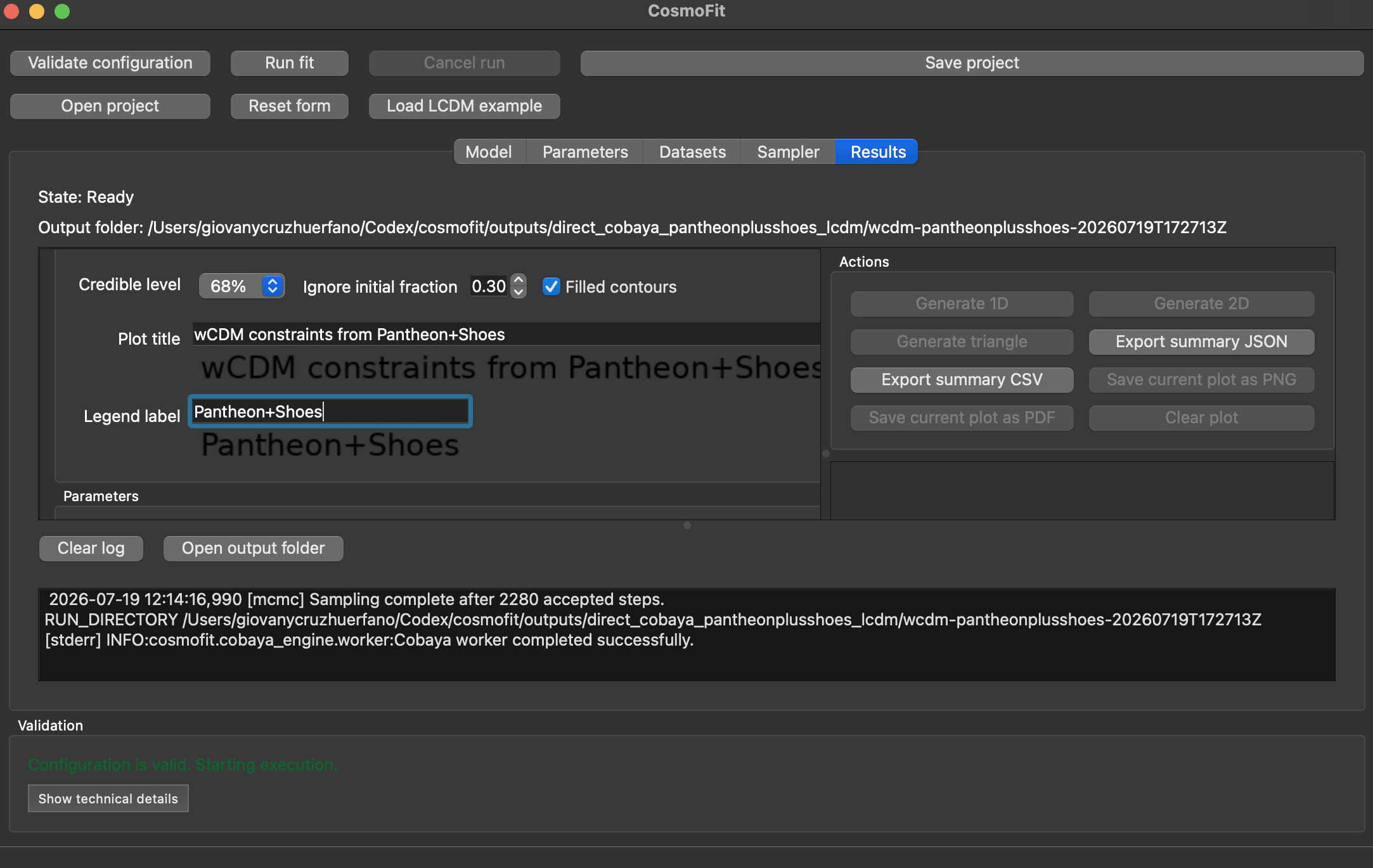}
    \caption{
        Posterior-processing configuration after completion of the run. The
        first \(30\%\) of the chain is ignored, the primary credible level is
        set to \(68\%\), and filled contours are enabled.
    }
    \label{fig:wcdm_getdist_configuration}
\end{figure*}

Figure~\ref{fig:wcdm_getdist_configuration} shows the corresponding
post-processing configuration. Filled contours were enabled, and the plot
title and legend were assigned before generating the graphical products.

\FloatBarrier
The user may select one parameter to obtain a one-dimensional marginalized
distribution, two parameters to generate joint contours, or several parameters
to produce a triangle plot. In the present example,
\(H_{0}\), \(\Omega_{m}\), and \(w\) were selected together.

\begin{figure*}[htbp]
    \centering
    \includegraphics[width=0.8\textwidth]{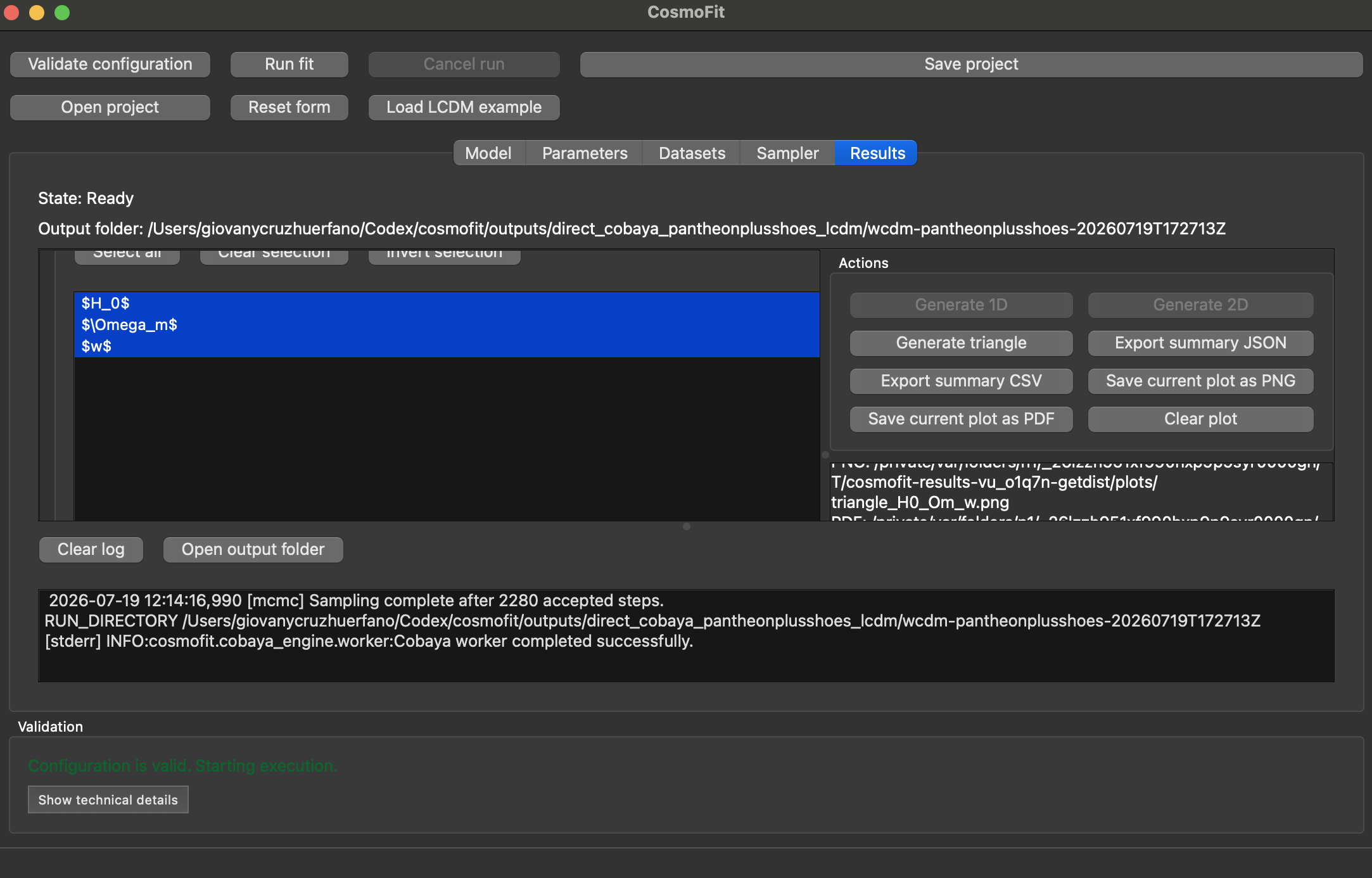}
    \caption{
        Selection of \(H_{0}\), \(\Omega_{m}\), and \(w\) for posterior
        visualization. The interface can generate one-dimensional,
        two-dimensional, and triangle plots, as well as export numerical
        summaries.
    }
    \label{fig:wcdm_plot_selection}
\end{figure*}
\FloatBarrier
Because posterior processing is independent of the sampling stage, the same
completed chains can be reanalyzed with different ignored fractions, credible
levels, parameter selections, or plotting options without repeating the MCMC
calculation.

\subsection{Numerical and graphical results}
\label{subsec:wcdm_results}

After removing the initial fraction, 1596 sample rows remained in the stored
chain. The marginalized numerical results are summarized in
Table~\ref{tab:wcdm_numerical_summary}.

\begin{table}[h]
    \centering
    \caption{
        Posterior summary for the illustrative spatially flat \(w\)CDM
        analysis with Pantheon+SH0ES.
    }
    \label{tab:wcdm_numerical_summary}
    \begin{tabular}{lccc}
        \hline
        Parameter & Mean & Standard deviation & Median \\
        \hline
        \(H_{0}\,[\mathrm{km\,s^{-1}\,Mpc^{-1}}]\)
        & \(73.4973\) & \(1.04405\) & \(73.4954\) \\
        \(\Omega_{m}\)
        & \(0.295457\) & \(0.0652577\) & \(0.299333\) \\
        \(w\)
        & \(-0.927074\) & \(0.143111\) & \(-0.919027\) \\
        \hline
    \end{tabular}
\end{table}

The same quantities are displayed directly in the results panel, together
with the retained sample count, number of loaded chains, ignored fraction, and
a preview of the generated triangle plot.

\begin{figure*}[htbp]
    \centering
    \includegraphics[width=0.8\textwidth]{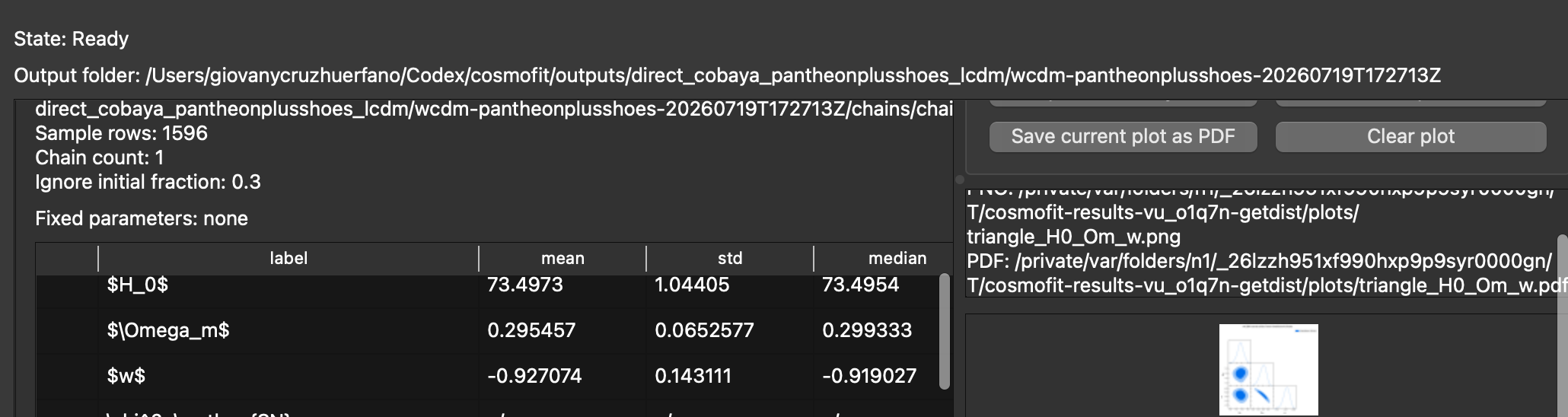}
    \caption{
        Numerical posterior summary displayed by \textsc{CosmoFit}. The
        panel reports the retained sample count and the marginalized mean,
        standard deviation, and median of \(H_{0}\), \(\Omega_{m}\), and
        \(w\).
    }
    \label{fig:wcdm_numerical_summary}
\end{figure*}
\begin{figure}[htbp]
    \centering
    \includegraphics[width=0.8\linewidth]{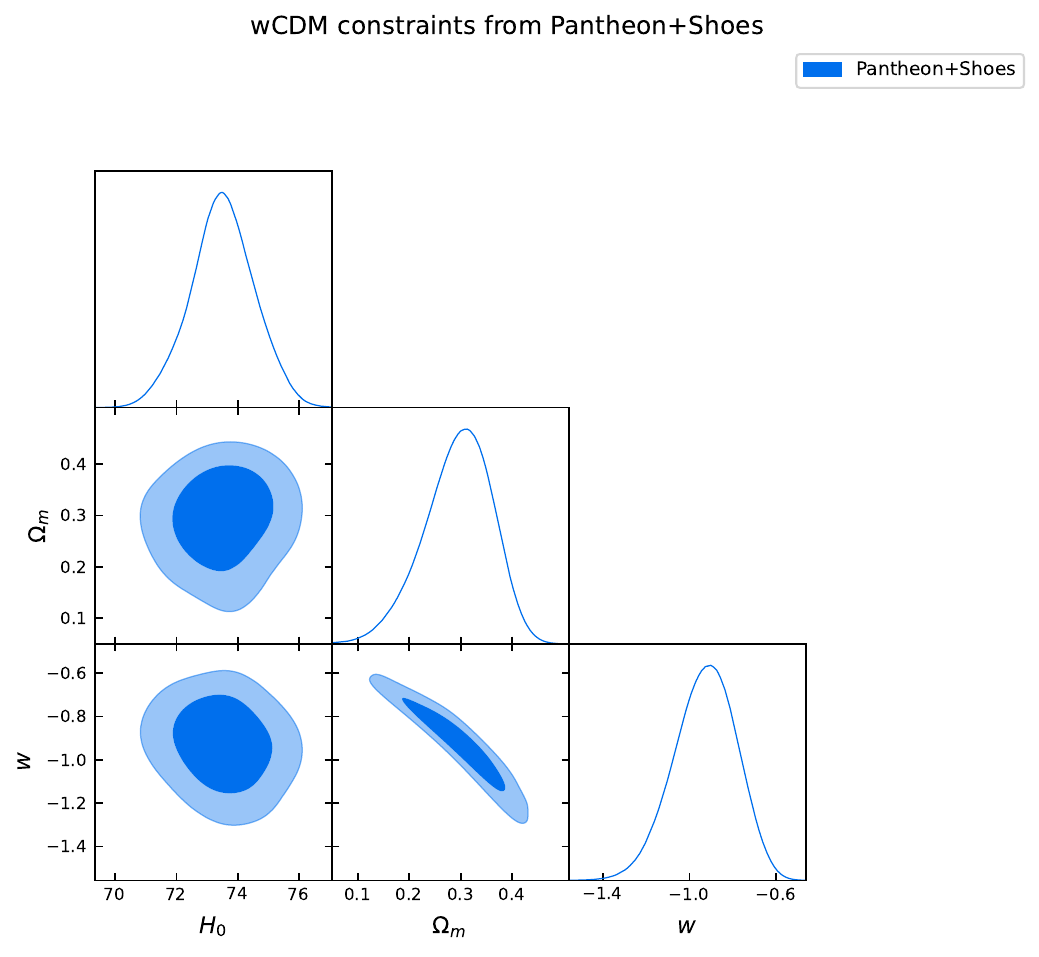}
    \caption{
        Triangle plot generated by \textsc{CosmoFit} for the spatially flat
        \(w\)CDM model with Pantheon+SH0ES. The diagonal panels show the
        marginalized distributions of \(H_{0}\), \(\Omega_{m}\), and \(w\),
        and the lower panels show their joint credible regions.
    }
    \label{fig:wcdm_triangle_plot}
\end{figure}

\FloatBarrier
The publication-quality triangle plot is shown in
Fig.~\ref{fig:wcdm_triangle_plot}. The diagonal panels contain the
one-dimensional marginalized distributions, while the lower panels show the
joint credible regions.

\FloatBarrier

The one-dimensional distributions are approximately unimodal, and the close
agreement between their means and medians indicates only mild asymmetry. The
most visible degeneracy appears in the \(\Omega_{m}\)--\(w\) plane, where the
elongated contour reflects the partial compensation between both parameters
in the luminosity-distance relation.

The marginalized value

\begin{equation}
    w=-0.927074\pm0.143111
\end{equation}

is compatible with the cosmological-constant value \(w=-1\), differing from it
by approximately \(0.51\) posterior standard deviations.

This result should be interpreted as an illustration of the complete
graphical workflow rather than as a definitive cosmological constraint. The
posterior summary was obtained from one stored chain, and the main purpose of
the example is to show that \textsc{CosmoFit} can connect model definition,
sampling, numerical summaries, and publication-quality posterior plots within
a single reproducible analysis.

The numerical summaries can be exported in JSON or CSV format, while the
plots can be saved as PNG or PDF. The PDF format is particularly suitable for
publication because it preserves vector text, axes, and contour boundaries.

\section{Discussion}
\label{sec:discussion}

The results presented in the previous sections show that
\textsc{CosmoFit} provides a practical graphical environment for defining,
running, and analyzing Bayesian studies of background cosmology. Its main
contribution is not a new sampling method or a new observational likelihood,
but the integration of established tools into a transparent and reproducible
workflow.
By relying on \textsc{Cobaya} for MCMC sampling and on \textsc{GetDist} for
posterior analysis, the application preserves the numerical procedures already
used in cosmological inference. The interface simplifies the preparation of
the analysis by allowing the user to specify the expansion history, parameters,
priors, likelihoods, and sampler settings without manually writing the complete
\textsc{Cobaya} configuration.
This approach is particularly useful when the background evolution can be
written explicitly as \(H(z)\). Standard and non-standard expansion histories
can be tested by changing the model expression and the corresponding parameter
definitions, while the generated configuration remains available for
inspection and independent execution. The interface therefore reduces
repetitive technical work without hiding the scientific structure of the
analysis.
The complete procedure can be summarized as

\begin{equation}
\begin{aligned}
    \text{theoretical model}
    &\longrightarrow
    \text{parameters and priors}
    \longrightarrow
    \text{likelihood}
    \\
    &\longrightarrow
    \text{sampling}
    \longrightarrow
    \text{posterior constraints}.
\end{aligned}
\label{eq:discussion_workflow}
\end{equation}

Presenting these stages separately makes the relation between the physical
model, the observational data, and the statistical assumptions easier to
follow. It also helps identify configuration errors before a numerical run is
started.
This organization may also be useful in academic settings. Students who are
beginning to work with cosmological inference can focus first on the physical
meaning of the model, priors, likelihoods, and posterior distributions without
being limited by an initial lack of experience with programming or
configuration files. At the same time, the generated inputs and outputs remain
accessible, so the interface can also serve as a transition toward direct
command-line or script-based analyses.
Transparency is therefore an essential part of the implementation. For each
run, \textsc{CosmoFit} preserves the normalized configuration, the generated
\textsc{Cobaya} input, metadata, logs, chains, and numerical summaries. The
graphical interface is not the only record of the analysis, and every relevant
choice can be inspected after the calculation has finished.
The restricted mathematical parser contributes to the same goal. It accepts
the redshift variable, declared parameters, supported operators, and a defined
set of mathematical functions. This provides enough flexibility for a broad
class of background models while keeping the model definition consistent with
the internal numerical representation.
The validation results indicate that the graphical layer preserves the
underlying inference problem. In the cosmic-chronometer comparison, the
posterior means obtained through \textsc{CosmoFit} agreed at the sub-percent
level with those from an independently configured \textsc{Cobaya} analysis.
The synthetic-recovery test also showed that the full workflow can recover
known fiducial parameters within the expected credible regions.
These checks are important because convenience alone is not sufficient for a
scientific interface. The agreement with independent calculations supports the
interpretation of \textsc{CosmoFit} as an orchestration layer over
\textsc{Cobaya}, rather than as a separate statistical implementation.
The illustrative \(w\)CDM example with Pantheon+SH0ES demonstrates the complete
workflow, from the definition and validation of \(H(z)\) to the generation of
numerical summaries and triangle plots. It also shows that the same completed
chains can be reprocessed with different plotting options and posterior
settings without repeating the MCMC calculation.
The present release is limited to background cosmology with spatially flat
distance relations. It does not yet compute cosmological perturbations or
connect directly to Boltzmann solvers such as \textsc{CLASS} or
\textsc{CAMB}. In addition, the model must currently be supplied as an
explicit function of redshift. Models defined only through coupled
differential equations would require an additional numerical background
solver.
The current implementation also supports a limited set of priors and
likelihoods. Future versions could include Gaussian, log-uniform, and
user-defined priors, together with additional background probes such as
baryon acoustic oscillations.
The modular separation among the graphical interface, expression parser,
configuration builder, likelihood layer, execution worker, and posterior tools
provides a natural basis for these extensions. New components can be added
without changing the overall workflow, provided that the required theoretical
quantities can be connected to the internal model representation.
Overall, \textsc{CosmoFit} provides an intermediate solution between a fully
manual inference pipeline and a closed black-box application. It reduces the
initial burden of programming and configuration while preserving the
scientific assumptions, numerical inputs, chains, logs, and posterior products
required for inspection and reproducibility.
\section{Conclusions}
\label{sec:conclusions}
We have presented \textsc{CosmoFit}, a desktop graphical interface designed
to facilitate Bayesian parameter inference for background cosmological
models. The application integrates model definition, parameter configuration,
observational-likelihood selection, MCMC execution, and posterior analysis
within a unified workflow based on \textsc{Cobaya} and \textsc{GetDist}.
The interface allows users to define an explicit expansion history \(H(z)\),
specify fixed and sampled parameters, assign prior intervals and proposal
scales, select supported observational datasets, and configure the MCMC
sampler without manually constructing the complete underlying configuration
files. The generated scientific inputs and outputs nevertheless remain
available, preserving the transparency and reproducibility of the analysis.
The implementation was verified through unit tests, a synthetic-parameter
recovery experiment, and a direct comparison with an independently configured
\textsc{Cobaya} analysis. The comparison showed sub-percent agreement between
the recovered posterior means, while the synthetic test confirmed that the
complete workflow can recover known fiducial parameters within the expected
credible regions. These results support the interpretation of
\textsc{CosmoFit} as a graphical orchestration layer that preserves the
underlying statistical calculation.
The illustrative flat \(w\)CDM analysis demonstrated the complete use of the
application, from the graphical definition of the Hubble function to the
generation of marginalized posterior distributions and numerical summaries.
The example also showed how the same completed chains can be processed with
different \textsc{GetDist} options and exported in graphical and tabular
formats.
The current release is focused on background cosmology with spatially flat
distance relations and models that can be expressed through an explicit
function \(H(z)\). Within this scope, \textsc{CosmoFit} provides a practical
environment for testing standard and non-standard expansion histories using
supernova and cosmic-chronometer data. Its modular architecture also provides
a basis for the future incorporation of additional priors, likelihoods,
background observables, and numerical model components.
Beyond its use in research workflows, the interface may also support academic
activities by reducing the initial programming barrier associated with
cosmological inference. This can allow users to focus first on the connection
between the theoretical model, observational data, Bayesian assumptions, and
posterior constraints, while retaining access to the generated files for more
advanced independent analyses.
Overall, \textsc{CosmoFit} offers an intermediate approach between direct
manual configuration and a closed black-box application. It simplifies the
construction and execution of background-level cosmological analyses while
preserving the inputs, logs, chains, and posterior products required for
inspection, validation, and reproducibility.
\section*{Code Availability}

The source code of \textsc{CosmoFit} v0.1.0, together with the installation
instructions, unit tests, validation scripts, and example configuration files,
is publicly available at
\url{https://github.com/GiovanyCruz/cosmofit}.

\section*{Data Availability Statement}

The synthetic cosmic-chronometer dataset generated for the parameter-recovery
test and the numerical results supporting the validation analysis are publicly
available in the \textsc{CosmoFit} repository at
\url{https://github.com/GiovanyCruz/cosmofit}, within the
\texttt{validation/} directory.

The Pantheon+, Pantheon+SH0ES, and Union3 data analyzed in this study are
secondary data products accessed through their official \textsc{Cobaya}
likelihood packages. Access to these data is governed by the distribution and
citation conditions of the original collaborations. The corresponding data
sources are cited in the reference list. No proprietary or restricted data
were generated or used in this study.

\begin{acknowledgments}
The author is grateful to Efraín Rojas for his valuable comments and
suggestions, which helped improve this work.
\end{acknowledgments}

\section*{Declarations}

\subsection*{Funding}

This work was supported by the Secretaría de Ciencia, Humanidades,
Tecnología e Innovación (SECIHTI), Mexico, through the program
\textit{Estancias Posdoctorales por México 2023(1)}, and was partially
supported by the Sistema Nacional de Investigadoras e Investigadores
(SNII), SECIHTI-Mexico.

\subsection*{Conflict of interest}

The author declares that he has no competing interests.

\bibliography{apssamp}

\end{document}